\documentclass[12pt]{article}
\def\fun#1#2{\lower3.6pt\vbox{\baselineskip0pt\lineskip.9pt
        \ialign{$\mathsurround=0pt#1\hfill##\hfil$\crcr#2\crcr\sim\crcr}}}

\renewcommand\({\left(}
\renewcommand\){\right)}

\newcommand\eq[1]{Eq.~(\ref{#1})}
\newcommand\eqs[2]{Eqs.~(\ref{#1}) and (\ref{#2})}

\newcommand\ee{\end{equation}}
\newcommand\be{\begin{equation}}
\newcommand\eea{\end{eqnarray}}
\newcommand\bea{\begin{eqnarray}}

\newcommand\TeV{\,\mbox{TeV}}
\newcommand\GeV{\,\mbox{GeV}}
\newcommand\MeV{\,\mbox{MeV}}

\newcommand\mpl{M_{\rm P}}

\newcommand{\lsim}{\mbox{\raisebox{-.9ex}{~$\stackrel{\mbox{$<$}}{\sim}$~}}}

\newcommand{\gsim}{\mbox{\raisebox{-.9ex}{~$\stackrel{\mbox{$>$}}{\sim}$~}}}

\def\dslash{\not{\hbox{\kern-2pt $\partial$}}}
\def\Dslash{\not{\hbox{\kern-4pt $D$}}}
\def\Oslash{\not{\hbox{\kern-4pt $O$}}}
\def\Qslash{\not{\hbox{\kern-4pt $Q$}}}
\def\pslash{\not{\hbox{\kern-2.3pt $p$}}}
\def\kslash{\not{\hbox{\kern-2.3pt $k$}}}
\def\qslash{\not{\hbox{\kern-2.3pt $q$}}}

 \newtoks\slashfraction
 \slashfraction={.13}
 \def\slash#1{\setbox0\hbox{$ #1 $}
 \setbox0\hbox to \the\slashfraction\wd0{\hss \box0}/\box0 }

\def\ee{\end{equation}}
\def\be{\begin{equation}}

\def\calp{{\cal P}}

\newcommand\sub[1]{_{\rm #1}}

\begin{document}

\date{}

\title{Can the curvaton paradigm accommodate a low inflation scale?}

\author{David H.~Lyth
\\
\\
Physics Department, Lancaster University, Lancaster LA1 4YB,  U.K.}
\sloppy
\maketitle

\begin{abstract}\noindent
The cosmological curvature perturbation may be generated  when  some
`curvaton' field, different from the inflaton, oscillates in
a background of unperturbed radiation. 
In its simplest form the curvaton  paradigm requires 
 the Hubble parameter during inflation to be bigger than $10^7\GeV$,
but this bound may be evaded if the curvaton field (or an associated
tachyon) is strongly coupled to a field which acquires a large value
at the end of inflation. 
As a result the curvaton paradigm might be useful in improving the
viability of low-scale
inflation models, in which  the 
supersymmetry-breaking mechanism is the same as 
the one  which operates in the vacuum.
\end{abstract}

\paragraph{Introduction}

In some of the most interesting inflation models, the inflationary 
potential comes from the same  SUSY-breaking mechanism that operates
in the vacuum, giving a Hubble parameter is of order the gravitino mass
\cite{treview}. 
Inventing some of the terminology, these are;
 (i) non-hybrid modular inflation \cite{modular,ks}, 
(ii) hybrid modular inflation \cite{rsg,running}, 
(iii) $\mu$-field  inflation invoking either
 gauge-mediated \cite{dr}, gravity-mediated \cite{steve1} or 
gaugino-mediated \cite{steve2} SUSY breaking, and
(iv) locked inflation \cite{dk}.

In gravity-mediated SUSY breaking the gravitino mass is of order 
$\TeV$, and in the other schemes it is some orders of magnitude lower
except for anomaly-mediated \cite{rs0} where it is up to $100\TeV$. 
An inflationary Hubble parameter of order the gravitino mass is therefore
low compared with the maximum value of order $10^{14}\GeV$ allowed by
the CMB anisotropy \cite{book}, and low compared with the value in other 
sensible-looking models of inflation \cite{treview}. 

One of the most important constraints on models of the very early Universe
is the existence of a   curvature perturbation $\zeta$, 
 known from observation to be present on cosmological scales
a few Hubble times before such scales start to enter the horizon.
At that  epoch, the earliest one at which it can be directly observed,
$\zeta$ is almost time-independent, with an almost scale-independent
spectrum $\calp_\zeta$ given by \cite{wmapspergel}
$\calp_\zeta^\frac12  \simeq 5\times 10^{-5}$.
This curvature perturbation is supposed to be generated by some field which 
is light during inflation, because indeed inflation converts 
 the vacuum fluctuation of
every such  field into an almost scale-invariant
 classical perturbation. 
The question is, which light field does the job? 

The usual answer is 
the inflaton \cite{book}.
 Unfortunately, this `inflaton
paradigm' tends to make life  difficult for the low-scale models
\cite{treview}.
In its  original  form,  non-hybrid modular inflation predicts
a curvature perturbation that is far too small, and so
do  the $\mu$-field  models,
unless one admits extreme  fine-tuning.\footnote
{The fine-tuning in 
model of \cite{steve1} might  be removed if the curvature perturbation is 
generated during preheating, from the decay products of the 
perturbed Higgs field \cite{steve2}. This is an alternative to the curvaton
mechanism that we are about to discuss.}
Hybrid modular inflation fares better, but some fine-tuning is still
required \cite{rsg} unless 
the inflaton mass is allowed to run \cite{running,ks} with the attendant
danger of a running of the spectral index  in conflict with observation.
Thus, it may reasonably be said that the inflaton paradigm makes life
difficult for low-scale inflation models.

According to the inflaton paradigm, the curvature perturbation has already
reached its observed value at the end of inflation and does not change
thereafter. The simplest alternative is to suppose that the curvature
perturbation is negligible at the end of inflation, being generated
later from the perturbation of some `curvaton'  field different from
the inflaton \cite{lw} (see also \cite{sylvia,lm}). 
This curvaton paradigm  has attracted a lot of attention [17--51,10]
because it opens up new possibilities both for observation and for
model-building.\footnote
{According to the scenario developed in the above papers, the curvature
perturbation is generated by the oscillation of the curvaton field.
A different idea \cite{decay} is that the field causing the 
curvature perturbation does so because its value determines the epoch
of reheating, and another is that it does so
through a  preheating mechanism \cite{steve2}. In all these cases one might
reasonably call the relevant field the curvaton, but the present paper
deals only with the original scenario.}

It is attractive to suppose \cite{dl} that the curvaton paradigm can 
be implemented in conjunction with  low-scale inflation models,
so as to liberate them from the troublesome requirement that the
inflaton  generate the curvature perturbation.
In this note I show that in the simplest version of the curvaton paradigm
this will not work,  because the curvaton can generate the observed curvature
perturbation only if the inflationary Hubble parameter exceeds $10^7\GeV$.
I go on to consider possible variants of the curvaton paradigm.

\paragraph{The simplest curvaton model}

In the simplest model \cite{lw}, the 
 curvaton field is  practically frozen, from the epoch of horizon
exit during inflation to the epoch when 
 the Hubble parameter $H$ falls below the curvaton mass $m$.
Also, the  curvaton potential in the early Universe 
is not appreciably modified, and in particular the mass $m$ is 
not modified.

With this setup,  
the value of the curvaton field $\sigma$ when the oscillation begins
is practically the same as its value $\sigma_*$ at the epoch
when the observable Universe leaves the horizon during inflation.
(Throughout, I will denote the latter  epoch by  star.)
The curvaton energy density is then
 $\rho_\sigma \sim m^2 \sigma_*^2$,
and the total energy density is $\rho\sim \mpl^2
m^2\sim \mpl^2 H^2$, making the  ratio 
\be
 \left. \frac{\rho_\sigma}{\rho}\right|_{H=m} \sim \frac{\sigma_*^2}{\mpl^2}
\label{ratio1}
\,.
\ee
This ratio is less than 1 by definition, corresponding to $\sigma_*\lsim \mpl$
which is a reasonable requirement. Afterwards it may grow, to achieve some
final value  which we denote by $r$. 
Such growth takes place during any era when
the non-curvaton energy density
is radiation-dominated. Let us assume for the moment that the growth is 
continuous, and  denote  the radiation density by
$\rho\sub r$. Discounting any variation in the effective number of species,
$\rho_\sigma/\rho\sub r$ is  proportional to the temperature, and
 curvaton decay increases this temperature by
a factor of order $(\rho/\rho\sub r)^\frac14$. 
(Complete thermalisation is assumed
after curvaton decay.) It follows \cite{dlnr} that
\be
\frac{r}{(1-r)^\frac34} \sim \frac{\sqrt{m\mpl}\sigma_*^2}{T\sub{dec}\mpl^2}
\,.
\ee
Remembering now that the growth may not actually 
be continuous we arrive at the 
inequality
\be
r \lsim \frac{\sqrt{m\mpl}\sigma_*^2}{T\sub{dec}\mpl^2}
\label{bound1}
\,,
\ee
which will be crucial in bounding $H_*$.

To obtain  rather precise results in a simple way, existing treatments
of the curvaton scenario assume that $\rho_\sigma/\rho$ does 
 grow significantly. We will use these results, while noting that our
rough order of magnitude estimates should be valid in the limiting case
where there is no growth. Once significant growth has taken place, 
the curvature perturbation is given by \cite{lw,luw}
\be
\zeta(t)  \simeq  \frac 13 \frac{\rho_\sigma}{\rho} 
\frac{\delta\rho_\sigma}{\rho_\sigma} 
\,;
\ee
In this expression   the  fractional curvaton 
 density  perturbation is evaluated on 
spatially flat slices of spacetime so that it is time-independent.
It may therefore be evaluated at the beginning of the oscillation,
when at each comoving point $\rho_\sigma$ is proportional
to $\sigma^2$, and to first order $\delta\rho_\sigma/\rho_\sigma
=2\delta\sigma/\sigma$.
After the curvaton decays $\zeta$ is supposed to remain constant 
until horizon entry, so that the  observed 
curvature perturbation is equal to the one just before curvaton decay,
\bea
\zeta &\simeq& \frac23 \delta\sigma/\sigma \label{zetaofsigma1}\\
&\simeq&  \frac23 \delta\sigma_*/\sigma_*
\,.
\label{zetaofsigma2}
\eea
Since the  spectrum of  $\delta\sigma_*$ is 
 $(H_*/2\pi)^2$, the spectrum of the observed 
curvature perturbation 
is therefore predicted to be   \cite{lw,luw}
\be
\calp_\zeta^\frac12 \simeq \frac {2r}{3} \frac{H_*}{2\pi\sigma_*}
\,.
\ee
Using the observed value $\calp_\zeta^\frac12 = 5\times 10^{-5}$
one finds that
\be
\sigma_* \simeq  \( 5\times 10^{-5} \times 3\pi \)^{-1} r  H_*
\label{hofsigma}
\,.
\ee 
Combining  \eqs{bound1}{hofsigma} leads to the bound
\cite{dlnr}
\be
\frac{\sqrt{m\mpl}H_*^2}{T\sub{dec}\mpl^2} \gsim 
\( 5\times 10^{-5} \times 3\pi \)^2
\label{bound2}
\,.
\ee

Imposing the 
BBN bound $T\sub{dec} > 1\MeV$ and the 
constraint $m< H_*$ gives   the advertised bound 
\be
H_* \gsim 10^7 \GeV
\label{hbound1}
\,.
\ee
Another bound comes from 
 the fact that the 
curvaton decay rate $\Gamma$ will be at least of order
$m^3/\mpl^2$, corresponding to gravitational-strength interactions.
Since the Hubble parameter at decay is of order $\Gamma$
this implies
\be
T\sub{dec}\sim \sqrt{\mpl\Gamma}\gsim \mpl (m/\mpl)^\frac32
\,,
\ee
and hence
\be
H_* \gsim 10^{11}\GeV \( \frac m {H_*} \)
\,.
\label{hbound2}
\ee
This is stronger than  \eq{hbound1} if $m \gsim 10\TeV$.

\paragraph{Evolution of the curvaton field}

As the simplest model is incompatible with low-scale inflation, we
 need to explore alternatives. 
One possibility is to 
 allow significant evolution of the curvaton field,
with the curvaton potential either unmodified in the early Universe,
or else altered only through a modification $\Delta m^2 \sim \pm
H_*^2$ of  the effective mass-squared that might be expected to come from
supergravity. (In the latter case, the actual modification should
be at least an order of magnitude or so below the expected one
during inflation, and preferably also afterwards \cite{dllr2}.)

Since we are dealing
with super-horizon scales, the evolution of the curvaton 
field at each comoving position is given 
by the same equation as for the unperturbed Universe,
\be
\ddot\sigma + 3H\dot\sigma + V'(\sigma) = 0
\,,
\label{sigeq}
\ee
with the initial condition $\sigma=\sigma_*$ and $\dot\sigma\simeq 0$.
Evaluated at the epoch $H=m$ this will give some value
$\sigma=g(\sigma_*)$, and a first-order   perturbation
\be
\delta \sigma \simeq  g' \delta\sigma_*
\label{firstorder} 
\,.
\ee

The effect of the evolution is to replace $\sigma_* $ by $g$
in \eqs{hbound1}{hofsigma}, 
and to multiply $H_*$ in  \eq{hofsigma} by the factor $g'$ corresponding
to the evolution  of $\delta\sigma$. 
The bound on $H_*$ is affected only 
by  the latter change, causing it to be multiplied 
by a factor $1/g'$. Unfortunately, the evolution  typically
 goes the wrong way, {\em decreasing} both the value and the
 perturbation of the curvaton \cite{dllr2}. The opposite can be true
if the evolving curvaton field almost reaches a maximum of its potential
\cite{dllr1},
but this happens only for a narrow range of $\sigma_*$,
which at least for the model of \cite{dllr1} 
can be achieved only if there has not
been too much inflation before our Universe leaves the horizon.

In addition to being difficult to achieve, a strong  increase in the
curvaton field brings with it the danger of generating too much
non-gaussianity. Indeed, extending \eqs{zetaofsigma1}{firstorder}
to second order  one finds that
the perturbation in the curvaton density when the oscillation begins 
is given by
\footnote
{In this and the preceding formulas one can proceed more rigorously
if the epoch $H=m$ is replaced by a somewhat later one, such that 
the harmonic oscillation of the curvaton field is well under way
and $\sigma$ is understood to be the amplitude of the oscillation
\cite{luw}.}
\be
 \frac{\delta \rho_\sigma}{\rho_\sigma} 
= 2 \frac{g'}{g} \delta\sigma_* + \(\frac{g'^2}{g^2} +\frac{g''}{g} \)
\(\delta\sigma_*\)^2
\ee
Repeating the argument in \cite{luw} (which implicitly assumed
$g''=0$) the non-linearity parameter becomes
\be
f\sub{NL} = \frac{5}{4r} \( 1 + \frac{gg''}{g'^2} \)
\,.
\ee
In the absence of evolution the 
current observational bound $f\sub{NL}\lsim 100$ is  achieved
for any $r\gsim 0.01$, but strong evolution might violate the bound
even for  $r=1$.
It would be worth checking  that  the bound is not violated
 for the  example of \cite{dllr2}.

\paragraph{The heavy curvaton}

The  curvaton may have an unsuppressed 
 coupling $\lambda\sigma^2\chi^2$ to some  field 
$\chi$, which is close to zero during inflation but moves
quickly to a large 
VEV at the end of inflation. (Candidates  for $\chi$ are the inflaton
field in the case of non-hybrid inflation,  and the  waterfall
field in the case of hybrid inflation.)
In that  case, the  effective  curvaton
mass-squared  increases by an amount $\lambda \langle \chi \rangle^2$
just after the end of inflation, allowing the true curvaton mass
to be bigger than $H_*$. One may call this kind of curvaton a
heavy curvaton.


The  heavy curvaton begins to oscillate as soon as its mass is
generated at  the end of inflation.
At this stage, its  energy density 
is of order $m^2 \sigma_*^2$ while the total density
is of order $\mpl^2 H_*^2$, so that \eq{ratio1} is replaced by
\be
 \left. \frac{\rho_\sigma}{\rho}\right|_{H=m} \sim 
\frac{m^2\sigma_*^2}{\mpl^2 H_*^2}
\,.
\label{ratio2}
\ee
This is less than 1 by definition, and  using \eq{hofsigma} this requires
\be
m/\mpl \lsim 5\times 10^{-4}/r < 5\times 10^{-2}
\label{mbound}
\,,
\ee
where the second inequality comes from the current \cite{wmapng}
non-gaussianity bound $r>.01$. 

Using \eq{hofsigma} and repeating the arguments
leading to 
\eqs{hbound1}{hbound2}, one  finds that 
 \eq{hbound1} is replaced by 
\be
H_*  \gsim \(10^7\GeV \)^5/m^4
\,,
\ee
while \eq{hbound2} is replaced by 
\be
H_* \gsim \( 10^{11}\GeV \)^2 /m
\,.
\ee
In the physical range $H_* <\mpl$ the second bound is always the stronger.
Imposing \eq{mbound} gives
\be
H_* \gsim (10^7\GeV)r \gsim 10^5\GeV
\,.
\ee
This is marginally compatible with low-scale inflation models, though it
will become incompatible if future non-gaussianity observations require
$r\simeq 1$.

\paragraph{Expansion of the curvaton field scale after inflation}

The last possibility that I consider applies only if the curvaton
is a PNGB corresponding to a symmetry which acts on the phase of
one or more complex fields \cite{lw,dlnr}. 
Taking the simplest case of a single complex
field $\Sigma$,
the potential will be of the form
\be
V(\Sigma) \simeq  V_0 - m_\Sigma^2 |\Sigma|^2 + \lambda \mpl^{-n}
|\Sigma|^{4+n} + \cdots
\,,
\ee
with $n\geq 0$.
The third term is the  term mainly responsible for the stabilization of
$|\Sigma|$, which gives it  a VEV 
\be
v\sim \( m_\Sigma^2\mpl^n/\lambda \)^\frac 1{2+n}
\,,
\ee
and defines  the curvaton field through $\Sigma = v\exp(i\sigma/\sqrt 2 v)$.
The dots  indicate higher powers of $\Sigma$, which in general are expected to
break the symmetry and generate the curvaton potential, 
as well as any quantum effects which do the same thing.

We  now suppose that there is a coupling $\lambda |\Sigma|^2 
\chi^2$ with {\em negative} $\lambda$, to a field $\chi$ which suddenly
 acquires
a large VEV after inflation. The negative coupling is un-typical, especially
in the context of supersymmetry, but it can be achieved as discussed for
instance in \cite{inverted}. 

With such a coupling,
the tachyonic mass $m_\Sigma$ will 
 suddenly increase at the end of inflation. This will 
suddenly increase the  VEV $v$  by some
factor, and increase  both $\sigma$  and 
$\delta \sigma$ by the same factor.
 As we saw earlier, the increase in $\delta\sigma$
 has the effect of
weakening the bound \eq{hbound1} on $H_*$,  which may 
allow low-scale inflation. Of course, there will in general also 
be a sudden change in the effective mass-squared
of the curvaton, by an amount which depends on the mechanism which 
explicitly 
breaks the global symmetry. This change may or may not cancel out the
beneficial effect of the increase in the value of the curvaton field
perturbation.

\paragraph{Conclusion}

In the  simplest form of the 
 curvaton paradigm, neither the curvaton field nor the form of the curvaton
potential change appreciably before the curvaton begins to oscillate.
This form    is incompatible with low-scale
inflation. 
On the other hand, the curvaton paradigm may become compatible with
low-scale inflation  if  the mass of the curvaton 
 increases sharply  at the 
end of inflation.
The same may be true if 
the curvaton field 
is the angular part of a complex field, whose tachyonic mass
increases sharply.

 \subsection*{Acknowledgments}
I thank my collaborators for continuing discussions about the curvaton
scenario, and in particular Yeinzon Rodriguez Garcia for comments on 
the initial version of this paper. This work was supported by PPARC grant
PPA/G/O/2000/00466.

 \newcommand\pl[3]{Phys.\ Lett.\ {\bf #1}  (#3) #2}
 \newcommand\np[3]{Nucl.\ Phys.\ {\bf #1}  (#3) #2}
 \newcommand\pr[3]{Phys.\ Rep.\ {\bf #1}  (#3) #2}
 \newcommand\prl[3]{Phys.\ Rev.\ Lett.\ {\bf #1}  (#3)  #2}
 \newcommand\prd[3]{Phys.\ Rev.\ D{\bf #1}  (#3) #2}
 \newcommand\ptp[3]{Prog.\ Theor.\ Phys.\ {\bf #1}  (#3)  #2 }
 \newcommand\rpp[3]{Rep.\ on Prog.\ in Phys.\ {\bf #1} (#3) #2}
 \newcommand\jhep[2]{JHEP #1 (#2)}
 \newcommand\grg[3]{Gen.\ Rel.\ Grav.\ {\bf #1}  (#3) #2}
 \newcommand\mnras[3]{MNRAS {\bf #1}   (#3) #2}
 \newcommand\apjl[3]{Astrophys.\ J.\ Lett.\ {\bf #1}  (#3) #2}

 \end{document}